%% file: eprint.tex
\documentclass[10pt, paper=a4, UKenglish]{article}
\usepackage{graphicx}
\def\Title#1{\begin{center} {\Large #1 } \end{center}}
\def\Author#1{\begin{center}{ \sc #1} \end{center}}
\def\Address#1{\begin{center}{ \it #1} \end{center}}

\newcommand\pubblock{\rightline{\begin{tabular}{l} Proceedings of the CTD 2023\\ \pubnumber\\
         \pubdate  \end{tabular}}}

\newenvironment{Abstract}{\begin{quotation} \begin{center} 
             \large ABSTRACT \end{center}\bigskip 
      \begin{center}\begin{large}}{\end{large}\end{center} \end{quotation}}

\newenvironment{Presented}{\begin{quotation} \begin{center} 
             PRESENTED AT\end{center}\bigskip 
      \begin{center}\begin{large}}{\end{large}\end{center} \end{quotation}}

\def\Acknowledgements{\bigskip  \bigskip \begin{center} \begin{large}
      \bf ACKNOWLEDGEMENTS \end{large}\end{center}}

\input econfmacros.tex

\usepackage[dvipsnames]{xcolor}
\usepackage{lineno}
\usepackage{subfig}
\usepackage{hyperref}
\usepackage{siunitx}

\newcommand\pubnumber{PROC-CTD2023-48}

\newcommand\pubdate{\today}

\def\affiliation{
On behalf of the Belle II Tracking and Vertexing Group, \\
Deutsches Elektronen Synchrotron (DESY), Germany}

\newcommand{\conference}{Connecting the Dots Workshop (CTD 2023)\\
October 10-13, 2023}

\usepackage{fancyhdr}
\definecolor{mygrey}{RGB}{105,105,105}
\begin{document}


\large
\begin{titlepage}
\pubblock

\vfill
\Title{Belle II track finding and hit filtering using precise timing information}
\vfill

\Author{Christian Wessel}
\Address{\affiliation}
\vfill

\begin{Abstract}
The SuperKEKB accelerator and the Belle II experiment constitute the second-generation asymmetric energy B-factory. SuperKEKB has recently set a new world record in instantaneous luminosity, which is anticipated to further increase during the upcoming run periods up to \SI{6e35}{\cm^{-2}\s^{-1}}. An increase in luminosity is challenging for the track finding as it comes at the cost of a significant increase of the number of background hits. The Belle II experiment aims at testing the Standard Model of particle physics and searching for new physics by performing precision measurements. To achieve these physics goals, including e.g. time-dependent measurements, the track finding and fitting has to deliver tracks with high precision and efficiency. As the track reconstruction is part of the online high level trigger system of Belle II there are also stringent requirements on the resource usage.

The Belle II tracking system consists of 2 layers of pixelated silicon detectors, 4 layers of double sided silicon strip detectors (SVD), and the central drift chamber. We will present the general performance and working of the track reconstruction algorithm of Belle II. In particular we will focus on the usage of hit time information from the silicon strip detector. The SVD has a very precise determination of the hit time, which will be used for the first time in the Belle II track finding in the next data taking period. These hit times are used for hit filtering, estimation of the time of collision, and the determination of the time of individual tracks. All of these are important tools to help to cope with the anticipated increase in background hits caused by the increase in luminosity.\end{Abstract}

\vfill

\begin{Presented}
\conference
\end{Presented}
\vfill
\end{titlepage}
\def\thefootnote{\fnsymbol{footnote}}
\setcounter{footnote}{0}
%

\normalsize 


\section{Introduction}
\label{intro}

High efficiency track finding with low contamination with fake tracks and tracks from machine-related background processes is a key requirement for the physics programme of the Belle II experiment.
While the performance of the existing Belle II tracking algorithms already provides tracks of high quality, further improvements are necessary to cope with the increase of background rate expected with the increase of instantaneous luminosity.
In this paper we describe the improvements to the Belle II tracking algorithms that will be applied in the upcoming data taking period starting in early 2024.
The article is structured as follows: Section \ref{sec:environment} describes the Belle II experiment and the experimental environment, Section \ref{sec:algorithms} briefly describes the tracking algorithms, Section \ref{sec:performance} gives an overview of the track finding performance, and Sections \ref{sec:svdtiming} and \ref{sec:flipandrefit} introduce new developments to reduce the number of tracks reconstructed from beam backgrounds and to reduce an observed charge asymmetry.

\section{Experimental environment at Belle II}
\label{sec:environment}

The Belle II experiment is located at the SuperKEKB $B$-factory at the KEK laboratory in Tsukuba, Japan.
SuperKEKB is an asymmetric accelerator which collides positrons and electrons at energies of \SI{4}{\GeV} and \SI{7}{\GeV}, respectively, resulting in a centre of mass energy of $\sqrt{s} = \SI{10.58}{\GeV}$ which corresponds to the mass of the $\Upsilon(4S)$ meson.
While the $\Upsilon(4S)$ meson itself decays into a pair of $B$ mesons, SuperKEKB also produces a significant amount of $\tau^+\tau^-$ and $q\bar q$ pairs.
While currently holding the world record instantaneous luminosity of $\mathcal{L} = \SI{4.7e34}{\cm^{-2}\s^{-1}}$, the aim is to achieve an instantaneous luminosity as high as $\mathcal{L} = \SI{6e35}{\cm^{-2}\s^{-1}}$ by the end of the decade.
The Belle II experiment so far collected $\int\mathcal{L}\text{d}t = \SI{424}{\femto\barn^{-1}}$ of data with the goal of collecting $\int\mathcal{L}\text{d}t = \SI{50}{\atto\barn^{-1}}$.
The high instantaneous luminosity is achieved by the so called nano beam scheme in which the vertical beam size of both beams at the interaction region is reduced to only \SI{50}{\nm}.

\subsection{The Belle II Detector}
\label{subsec:belleii}

Belle II is a multi-purpose near-4$\pi$ detector with dimensions of roughly $7 \times 7 \times 7 \, \text{m}^3$.
At the heart of Belle II is the vertex detector (VXD) comprising two layers pixel detector (PXD) based on DEPFET \cite{depfet} technology and four layers of double sided silicon strip detectors (SVD) \cite{svd}.
The PXD consists \num{7.68} million pixels at radii of \SIlist{14;22}{\milli\metre} with pixel sizes of \SI{50}{\micro\m} in $r-\varphi$ direction and between \SIlist{55;85}{\micro\m} in $z$ direction which is defined by the symmetry axis of the magnetic field and pointing towards the direction of the electron beam.
Its integration window has a length of \SI{20}{\micro\s} and the material budget is only 0.2\% radiation length per layer.

The SVD has an inner radius of \SI{39}{\milli\m} and an outer radius of \SI{135}{\milli\m} and consists of about 220 thousand strips with pitches of \SIrange{50}{75}{\micro\m} in $r-\varphi$ direction and \SIrange{160}{240}{\micro\m} in $z$ direction while each layer has a material budget of 0.7\% radiation length.
The forward part of the three outer layers is slanted with respect to the beam axis, in order to reduce the material budget seen by forward tracks.
The VXD is surrounded by the Central Drift Chamber (CDC).
It comprises more than 14 thousand sense wires with a maximum radius of \SI{112}{\cm} and drift cell sizes between \SIlist{1;2}{\cm}.
The wires are arranged in nine super layers, five of which are axial wires with the wire direction aligned with the global $z$ axis, and four stereo layers with a small angle w.r.t. the $z$ axis.
The whole tracking volume covers 2$\pi$ in $\varphi$ and coverage of the polar angle is $\ang{17} < \theta < \ang{150}$.
Particle identification detectors are located outside of the CDC: a time of propagation counter in the barrel region and an aerogel ring imaging Cherenkov detector in the forward region.
All of the above detectors are enclosed by the electromagnetic calorimeter, and the barrel part of which is itself surrounded by the superconducting solenoid which provides a nearly homogeneous magnetic field of \SI{1.5}{\tesla} along the $z$ axis.
The outermost detector is the $K_L^0$ and muon detector, which iron plates serve as showering material as well as return jokes for the magnetic field.

\subsection{Experimental environment}
\label{subsec:environment}

At $B$-factories, the experimental environment is very clean.
In particular, in hadronic events there are on average around 10 tracks per event with a soft momentum spectrum with the mean of the transverse momentum ($p_t$) distribution being at around \SI{400}{\MeV/c}.
Thus, multiple scattering is a potential issue in Belle II, hence the requirement of low material budget detectors described in Section \ref{subsec:belleii}.
Due to the soft momentum spectrum the tracking algorithms must provide high track finding efficiency for transverse momenta down to \SI{50}{\MeV/c} with a low rate of fake tracks.
At the $\Upsilon(4S)$ centre of mass energy, the Bhabha scattering and two-photon process cross sections are larger than the hadronic ones.
Since at SuperKEKB the bunch crossing occurs every  \SI{4}{\ns}, there is a significant number of tracks created in $e^+e^-$ collisions that are not the triggered one, that are reconstructed in the triggered event.
These tracks are real, but do not belong to the triggered event, so should be discarded in physics analysis.
At nominal luminosity, we expect only about 0.1\% of the hits in the PXD and about 10\% of the hits in the SVD per event are hits from signal processes with all the rest stemming from background processes \cite{background}.
Since the SVD provides precise hit time information with a resolution of less than \SI{4}{\ns} we will use them to reduce the number of beam background hits as well as fake hits in the future, with the goal of rejecting actual fake tracks as well as background tracks in the triggered events.

\section{Tracking algorithms in Belle II}
\label{sec:algorithms}

\begin{figure}[!htb]
  \centering
  \includegraphics[width=0.6\linewidth]{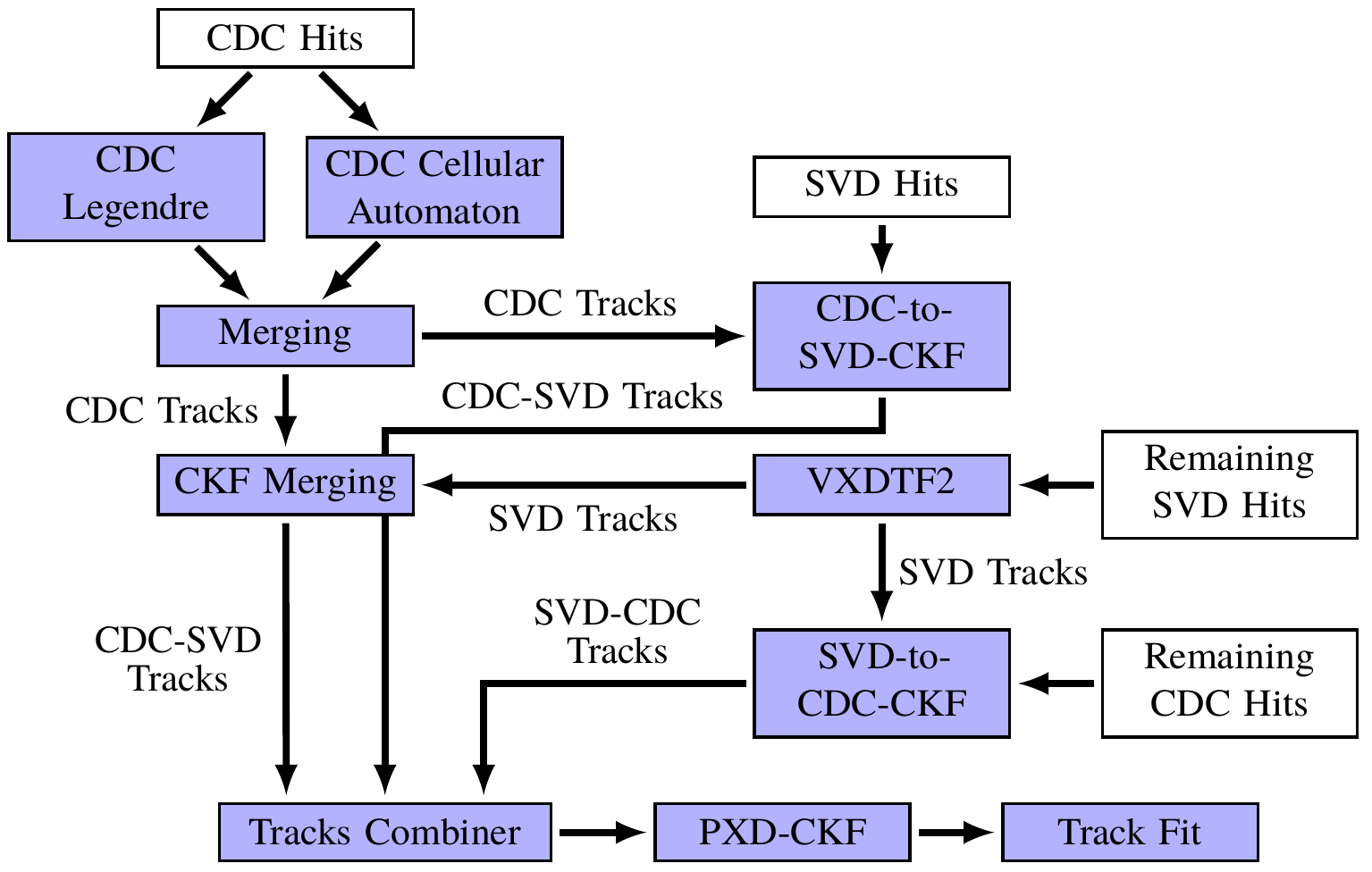}
  \caption{Tracking flow in Belle II.}
  \label{fig:trackingflow}
\end{figure}

In Belle II several different tracking algorithms are employed to account for the differences in detector technologies and they are described in more detail in \cite{b2tracking}.
In general, the GenFit2 library \cite{genfit} is used for track fitting and related tasks like track extrapolations.
All algorithms are implemented within the Belle II software and analysis framework basf2 \cite{basf2}.
The tracking chain is depicted in Figure \ref{fig:trackingflow}.
First, CDC standalone tracks are build using both a local and a global algorithm.
These tracks are then extrapolated into the SVD volume using a Combinatorial Kalman Filter (CKF, \cite{CKF}) to attach single SVD hits (CDC-to-SVD-CKF).
All unused SVD hits after this step are then used in the SVD standalone tracking algorithm (Vertex Detector Track Finder 2, VXDTF2) based on the sectors-on-sensors concept \cite{VXDTF}.
This algorithm internally employs geometrical as well as cuts on time difference between SVD hits to only accept combinations that can possibly be part of a track from the interaction region.
Afterwards, these SVD standalone tracks and the CDC tracks without any SVD hits attached are extrapolated to the CDC inner wall and combined if possible by another CKF (CKF merging).
Only the remaining SVD standalone tracks without any CDC hits attached to them are extrapolated towards the CDC with a third CKF (SVD-to-CDC-CKF) to attach CDC hits not consumed in the first CDC standalone track finding step.
After this step, all tracks are collected and extrapolated to the PXD, followed by the final track fit.
This procedure means that the PXD is not part of the actual track finding, i.e. we do not attempt to reconstruct tracks from PXD and SVD hits directly but rely on the existence of tracks before the PXD extrapolation.
The final track fit uses the Deterministic Annealing Filter (DAF) from GenFit2.

\section{Tracking performance}
\label{sec:performance}

The tracking performance is measured with a few figures of merit, the most important are the track finding efficiency, fake rate, and the clone rate.
A track is considered \textit{found} if its hit purity is larger than 66.7\% and its hit efficiency is larger than 5\%, and if the track fit for at least one mass hypothesis is converged.
Hit purity is defined as the ratio of true hits in a track and the total number of hits within that track, while the hit efficiency is defined as the fraction of correctly found hits of a simulated charged particle and the total number of hits of that particle.
If multiple reconstructed tracks exist for a given simulated particle and all of them fulfill both the hit purity and hit efficiency criteria, the one with the highest hit purity is considered the correct track and the other tracks are considered as \textit{clone} tracks.
If a reconstructed track does not fulfill the hit purity or the hit efficiency criteria it is considered to be a \textit{fake} track.
This, however, means that all true tracks from beam backgrounds are considered to be fake in simulation as no corresponding information from a simulated particle is available.
Thus, the reported fake rate consists of two parts: actual tracks from beam backgrounds and tracks made up from random combinations of hits.
Simulated beam background samples are based on simulations of the full accelerator using the SAD \cite{SAD} and Geant4 \cite{Geant4} for a given luminosity and scaled to the target luminosity based on dedicated beam background studies.
The results presented in this section are based on 300,000 simulated $B\bar B$ events using EvtGen \cite{EvtGen} with simulated backgrounds for an instantaneous luminosity of $\mathcal{L} = \SI{6e35}{\cm^{-2}\s^{-1}}$.

\begin{figure}[!htb]
  \centering
  \subfloat[]{
    \includegraphics[width=0.43\linewidth]{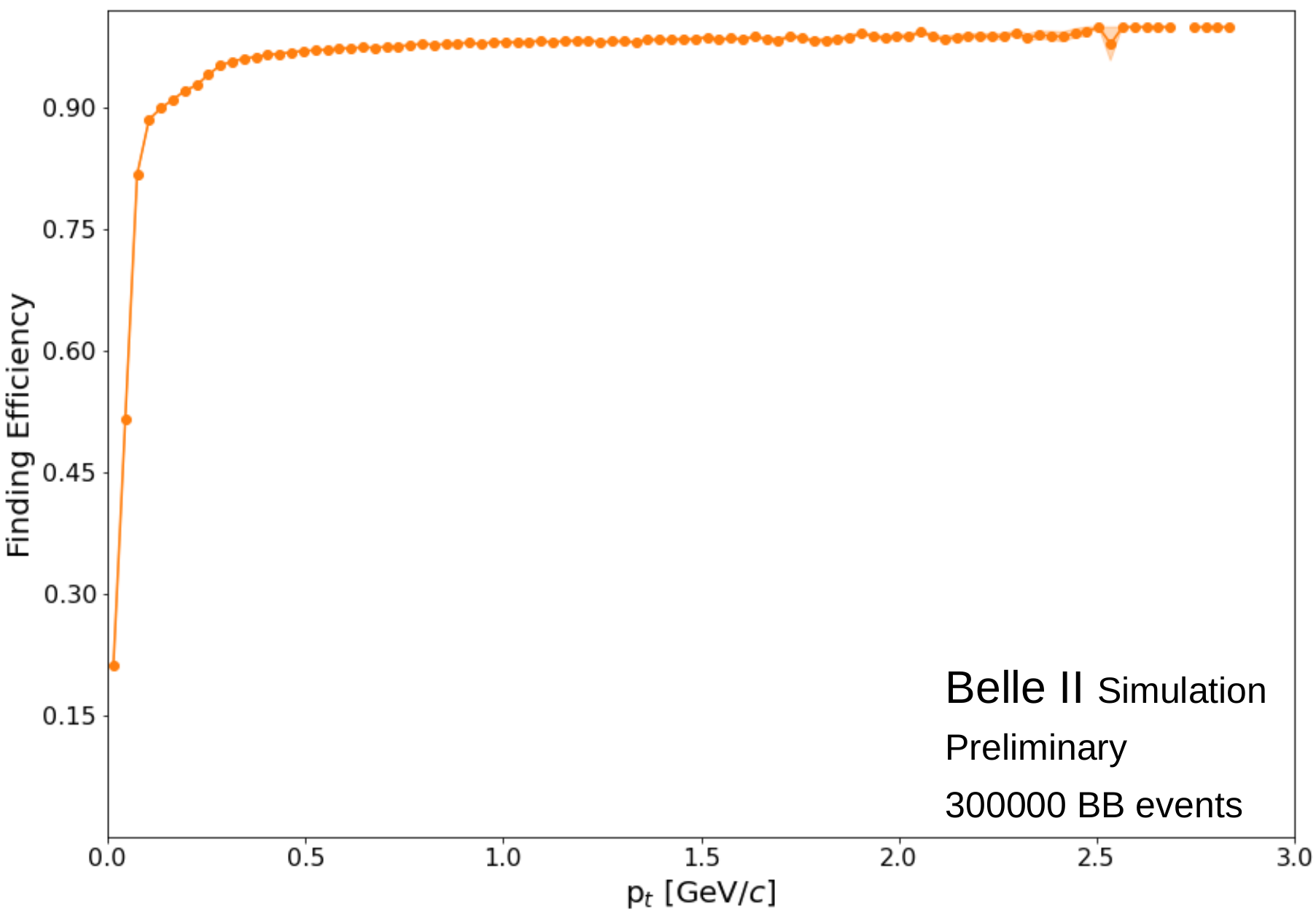}
    \label{fig:efficiencypt}
  }
  \qquad
  \subfloat[]{
    \includegraphics[width=0.43\linewidth]{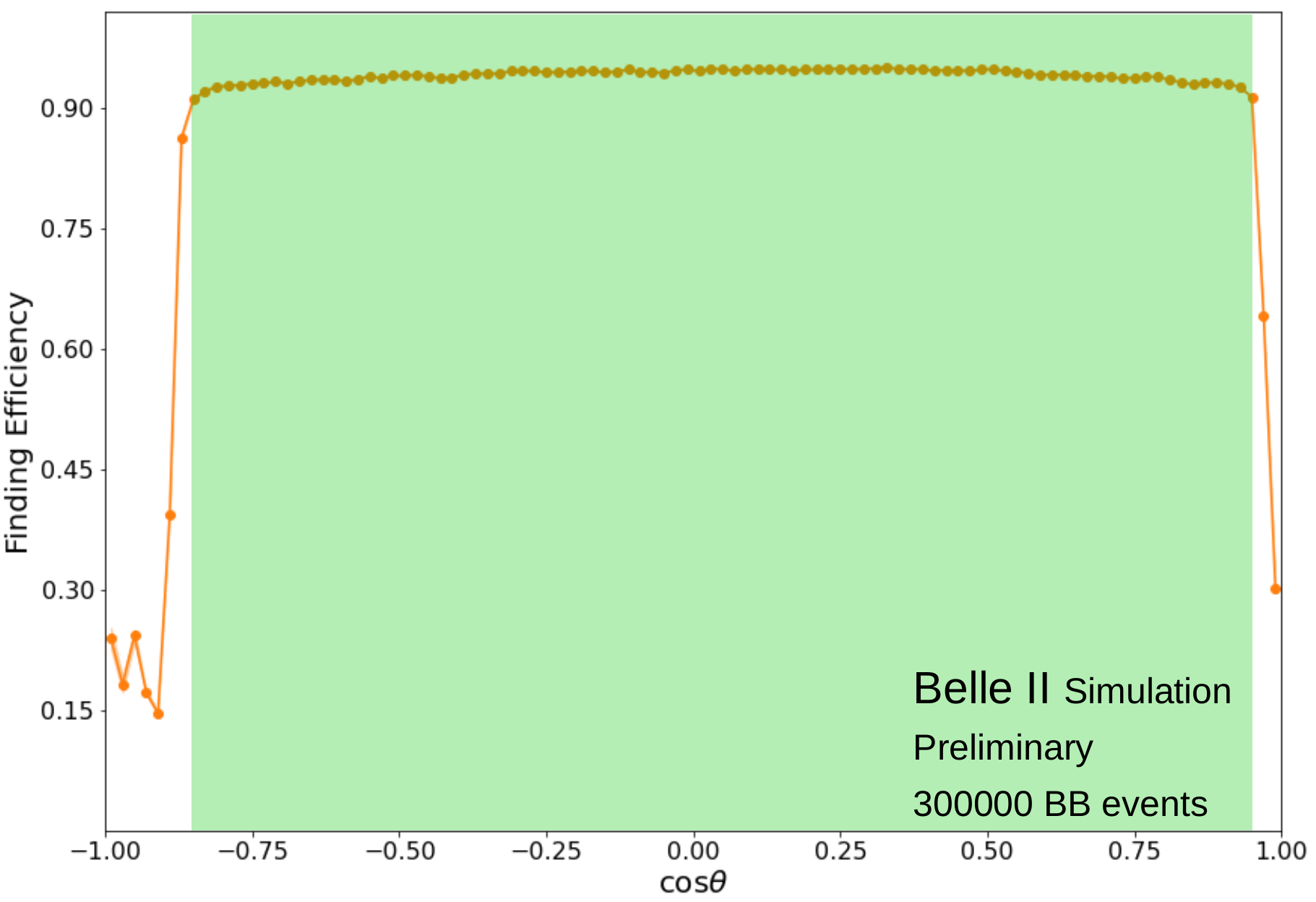}
    \label{fig:efficiencytheta}
  }
  \caption{Track finding efficiency as function of $p_t$ (a) and $\cos\theta$ (b). The green region in (b) marks the fiducial region of the Belle II tracking system.}
  \label{fig:efficiency}
\end{figure}

Figure \ref{fig:efficiencypt} shows the track finding efficiency as function of transverse momentum $p_t$.
A steep rise in efficiency is visible at low values of $p_t$ and the efficiency reaches 90\% at $p_t$ of around \SI{100}{\MeV/c}.
A small shoulder is visible at around $p_t$ = \SI{250}{\MeV/c}.
At $p_t$ values larger than this, particles do traverse the full CDC without curling back.
The efficiency keeps slightly increasing until the highest $p_t$ values in Belle II, with an average track finding efficiency of 93.6\% for generic $B\bar B$ events.
Figure \ref{fig:efficiencytheta} shows the track finding efficiency as function of $\cos\theta$ with $\theta$ being the polar angle.
The efficiency is almost flat within the fiducial region and falls off steeply outside as expected.

\begin{figure}[!htb]
  \centering
  \subfloat[]{
    \includegraphics[width=0.43\linewidth]{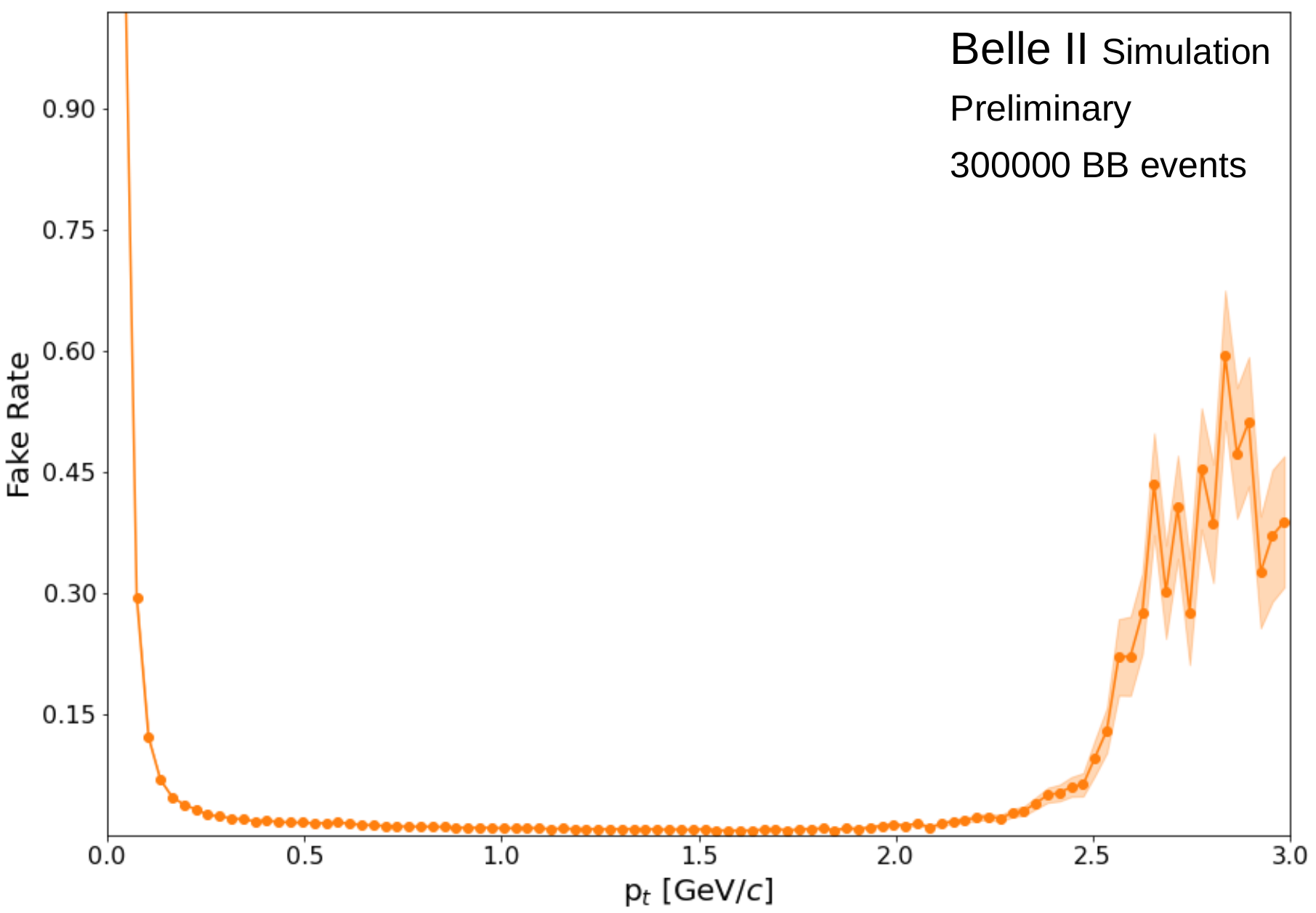}
    \label{fig:fakeratept}
  }
  \qquad
  \subfloat[]{
    \includegraphics[width=0.43\linewidth]{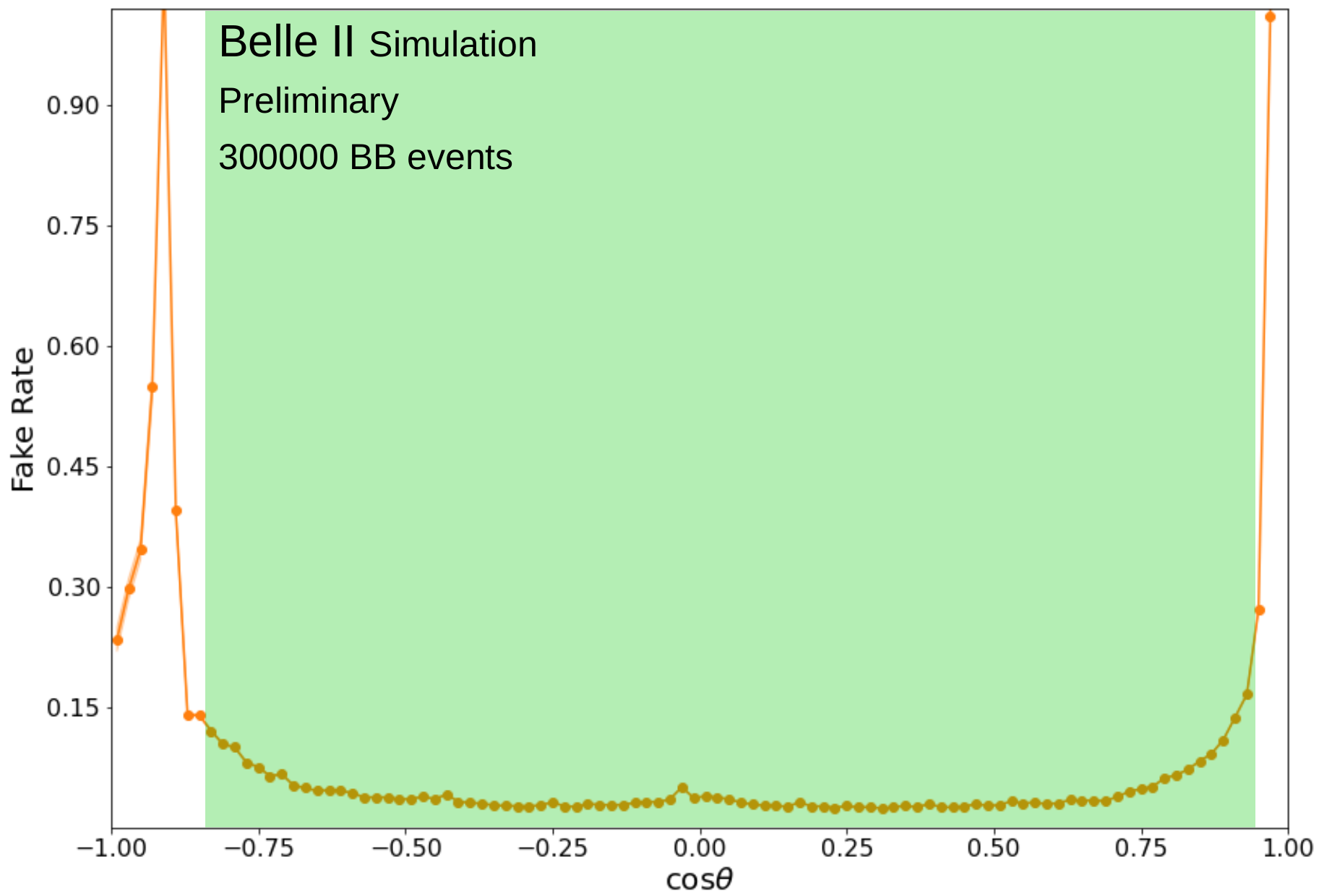}
    \label{fig:fakeratetheta}
  }
  \caption{Fake rate as function of $p_t$ (a) and $\cos\theta$ (b). The green region in (b) marks the fiducial region of the Belle II tracking system.}
  \label{fig:fakerate}
\end{figure}

Figure \ref{fig:fakeratept} shows the fake rate as function of transverse momentum.
While being close to zero over most of the $p_t$ range, a steep increase is visible at low values of $p_t$.
This is mostly caused by real tracks from small angle Bhabha scattering as well as low momentum tracks from the two photon process $e^+e^- \to e^+e^-\gamma\gamma \to e^+e^- f\bar f$ where the newly created fermion pair $f\bar f$ results in additional low momentum particles.
Most of these tracks are in the very forward and backward directions of the detector, as shown in Figure \ref{fig:fakeratetheta} where the fake rate increases close to the boarder of the fiducial region.
At high $p_t$ values, above \SI{2}{\GeV/c}, tracks are likely build from random combinations of hits, and there are beam background tracks from large angle scattering from $e^+e^- \to e^+e^-(\gamma)$ or $e^+e^- \to \mu^+\mu^-(\gamma)$.
In addition, due to the soft momentum spectrum, only very few real tracks from $B\bar B$ events are present in this $p_t$ region, leading to an increased fake rate.

\section{Exploit the precise SVD time}
\label{sec:svdtiming}

\begin{figure}[!htb]
  \centering
  \subfloat[]{
    \includegraphics[width=0.41\linewidth]{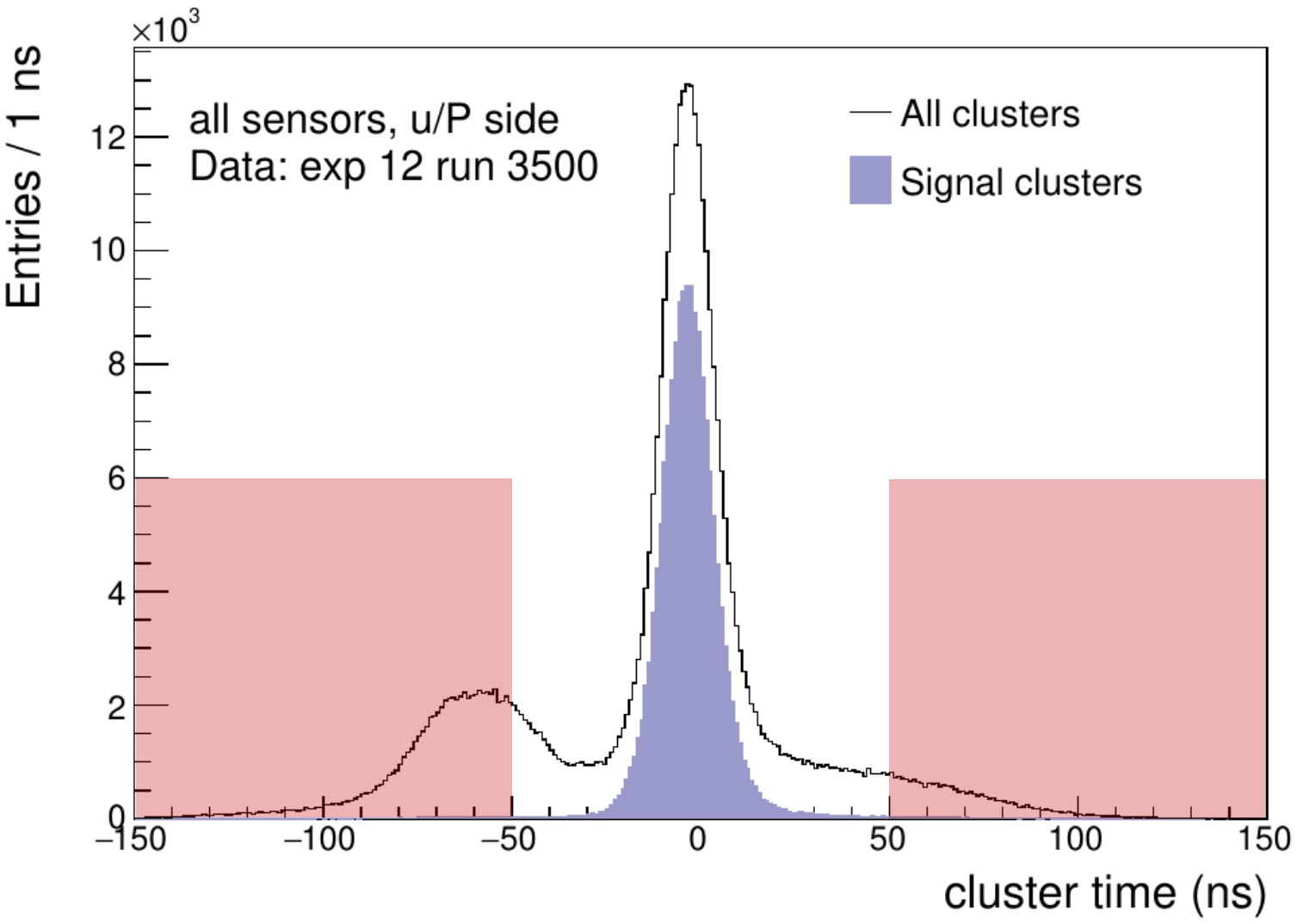}
    \label{fig:SVDabstimecuts}
  }
  \qquad
  \subfloat[]{
    \includegraphics[width=0.51\linewidth]{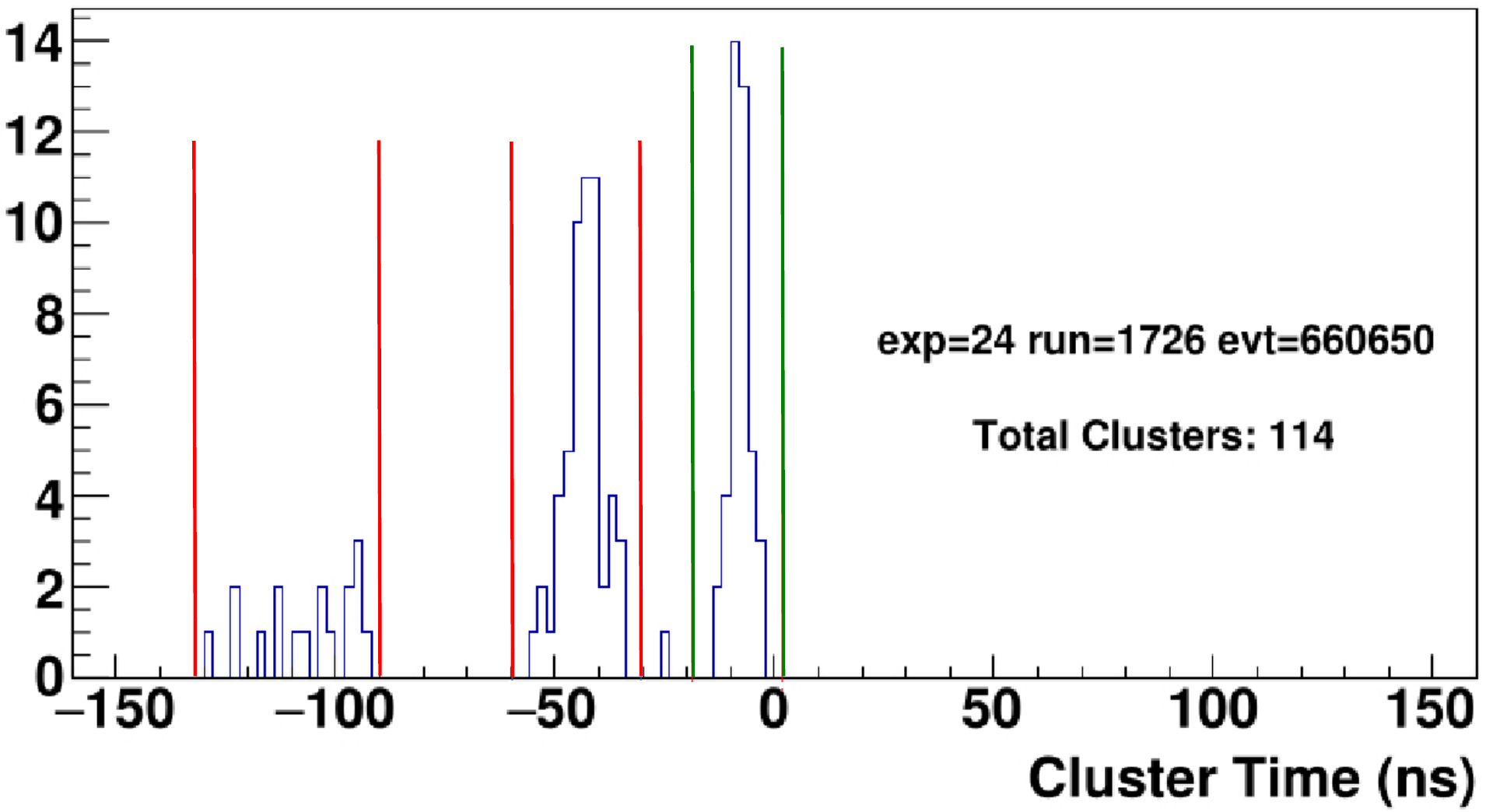}
    \label{fig:SVDgrouping}
  }
  \caption{Absolute cuts on SVD cluster time distribution, the red areas show the clusters rejected by the cut on absolute time (a) and grouping of SVD clusters based on their reconstructed time (b).
           A large fraction of background clusters remains after the cuts on the absolute cluster time of \SI{50}{\ns}.
           In (b) the cluster time group close to 0, enclosed with green bars, is the priority group, while the other clusters are assigned a higher group ID.
           Depicted is the distribution of SVD cluster times for only one event.}
  \label{fig:svdtiming}
\end{figure}

The SVD provides a precise measurement of the hit time, with a resolution better than \SI{4}{\ns} on data.
Details on how the cluster time are computed can be found in \cite{svd}.
Figure \ref{fig:svdtiming} shows the distribution of the cluster time. The signal (on-time) clusters populate the peak at $t=0$.
The beam-background hits are created uniformly distributed in time\footnote{SuperKEKB bunch crossing frequency is \SI{256}{\MHz}.}, we observe a bump at around \SI{-60}{\ns} due the accumulation of hits produced on the detector before the beginning of the integration window corresponding to the trigger, while the tail on the right is cut because of the asymmetry of the shaped response of the APV25, the readout chip.

The clusters attached on the outgoing arm of the tracks are used to compute the absolute track time.
Tracks with transverse momentum larger than \SI{250}{\MeV/c} are then used to compute the time of the $e^+e^-$ collision (EventT0), and the track time is then corrected with this information, in order to provide analysts with the time at which the track is created with respect to the $e^+e^-$ collision.
Preliminary studies have shown that rejecting tracks more than \SI{20}{\ns} far from the collision will significantly reduce the fake rate, up to 20\% in high background conditions.
Note that the trigger jitter, i.e. the width of the EvenT0 distribution, is $\approx$ \SI{10}{\ns}.
In the past, the EventT0 was estimated using CDC hit information.
This approach involved several passes of the GenFit2 DAF and thus had a long execution time.
For the upcoming data taking periods this algorithm was superseded by the determination of EventT0 using SVD information as described above that is approximately 2000 times faster in execution time and provides similar resolution of the order of \SI{1}{\ns}.
If no tracks with SVD hits associated to them are present in an event, the EventT0 is estimated using calorimeter information alone.\footnote{This situation only occurs in less than 1\% of all hadronic events.}

The precise timing information of SVD hits can be exploited further.
Previously, only absolute cuts on the SVD hit times were applied in simulation for the nominal luminosity, only accepting those hits with an absolute cluster time $|t| < \SI{50}{\ns}$ and the relative cluster times within a space point consisting of two clusters with orthogonal strip orientations $u$ ($r-\varphi$) and $v$ ($z$) of $|t_u - t_v| < \SI{20}{\ns}$.
These cuts were selected analysing data integrated over full runs, and a significant amount of background hits remained after these cuts, as shown in Figure \ref{fig:SVDabstimecuts}, and their number will only increase with increasing luminosity.

A new concept was recently developed based on SVD cluster time distributions \textit{per event}.
Clusters from the same bunch crossing are close in time and form groups of cluster times, cf. Figure \ref{fig:SVDgrouping}.
First, the individual SVD cluster times are filled in a histogram and a priority group is defined as the largest group of bins closest to 0.
The group is fitted with a Gaussian, and additional bins are attempted to be added if it is less than seven standard deviations away, after which the Gaussian fit of the group is repeated.
If a cluster time is more than seven standard deviations away from a group it is considered not belonging to a given group.
The procedure is then repeated with the other local maxima in the histogram until all SVD cluster times are added to at least one group.
Finally, only the clusters from the priority group are used to create 3D space points.
This method significantly reduces the number of off-time clusters used in an event while keeping more than 99.8\% of signal clusters.
Eventually, this leads to a reduction of the number of reconstructed background tracks and thus reduces the number of tracks considered as fakes.

\begin{figure}[!htb]
  \centering
  \includegraphics[width=0.5\linewidth]{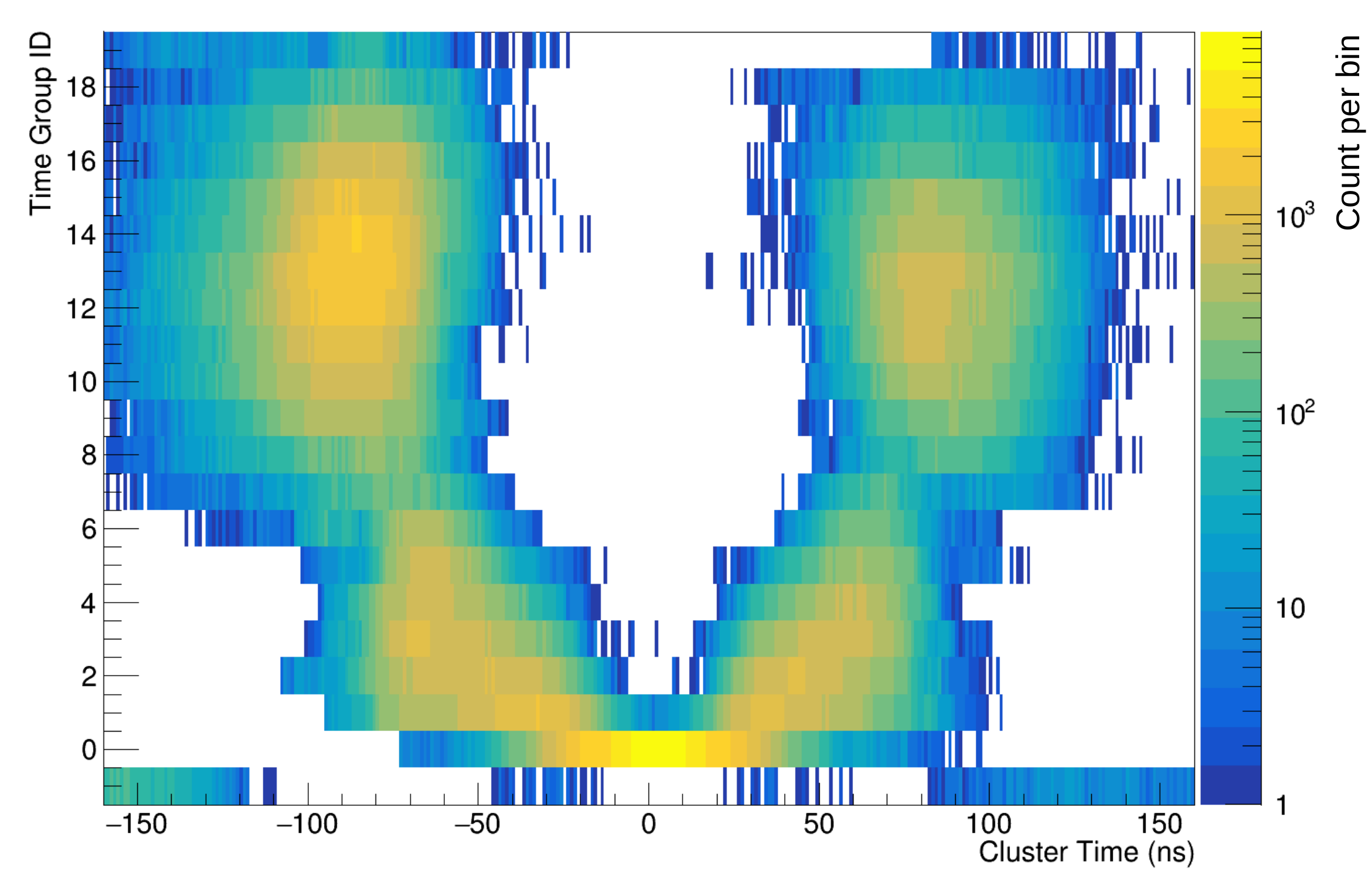}
  \caption{Cluster time grouping. The priority cluster group with ID = 0 is mostly free of background, while background clusters have group IDs $\neq$ 1 with their cluster times usually differing significantly from \SI{0}{\ns}.}
  \label{fig:SVDgrouping2D}
\end{figure}

An example for the performance of this time grouping is shown in Figure \ref{fig:SVDgrouping2D}.
The priority group corresponds to group ID 0 and it contains basically only signal clusters.
Background clusters have higher group IDs, and up to 30 groups are identified per event in the studies conducted thus far.

\begin{table}[!htb]
  \begin{center}
    \caption{Overview of improvements in tracking performance due to SVD cluster time grouping. A significant decrease of the fake rate is seen.}
    \label{tab:SVDgrouping}
    \begin{tabular}{l|ccc}
      \hline
                                & Hit time grouping off & Hit time grouping on  & Relative difference \\
      \hline
      Track finding efficiency  & 93.67 $\pm$ 0.24\%   & 93.69 $\pm$ 0.24\% & +0.02\% \\
      Fake rate                 &  9.55 $\pm$ 0.29\%   &  4.37 $\pm$ 0.20\% & \textcolor{Green}{$-54\%$} \\
      Clone rate                &  3.81 $\pm$ 0.19\%   &  3.56 $\pm$ 0.18\% & \textcolor{Green}{$-6.6\%$} \\
      \hline
    \end{tabular}
  \end{center}
\end{table}

The results of SVD cluster time grouping are summarised in Table \ref{tab:SVDgrouping}.
No degradation in track finding efficiency is visible, but the fake rate is reduced significantly from 9.55\% to 4.37\% which corresponds to a reduction of more than 50\%.
In addition, the clone rate also is slightly reduced as in some cases the track times of the curling parts are large enough to belong to a different group than the priority group, thus the space points of these tracks are not considered in the first place.

\section{Track flip and refit to improve charge reconstruction}
\label{sec:flipandrefit}

\begin{figure}[!htb]
  \centering
  \subfloat[]{
    \includegraphics[width=0.43\linewidth]{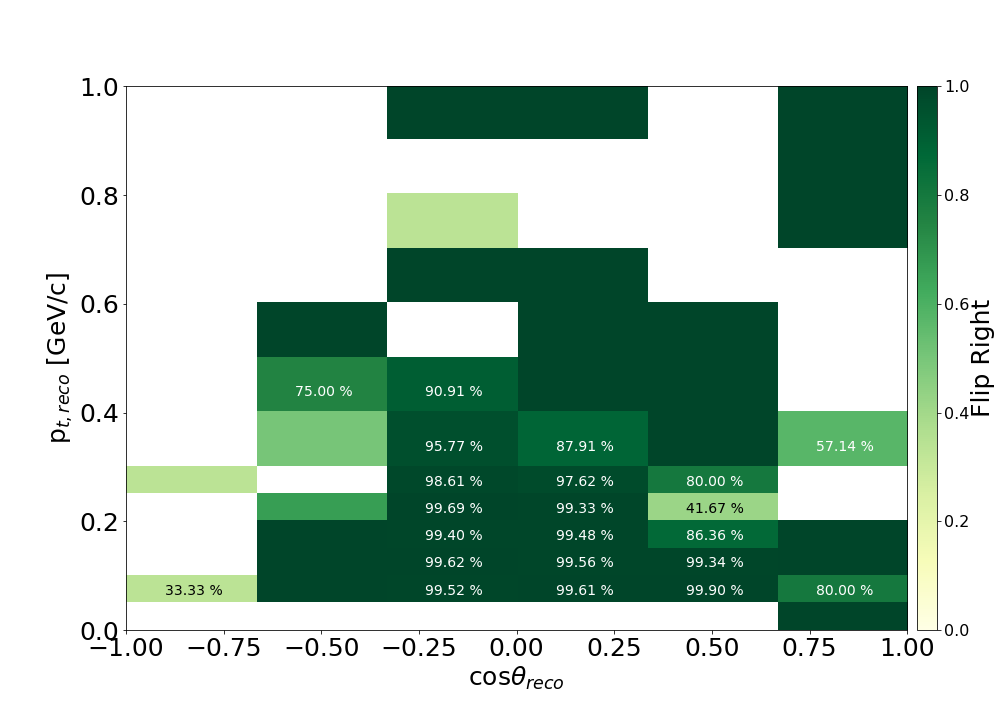}
    \label{fig:CorrectlyFlipped}
  }
  \qquad
  \subfloat[]{
    \includegraphics[width=0.43\linewidth]{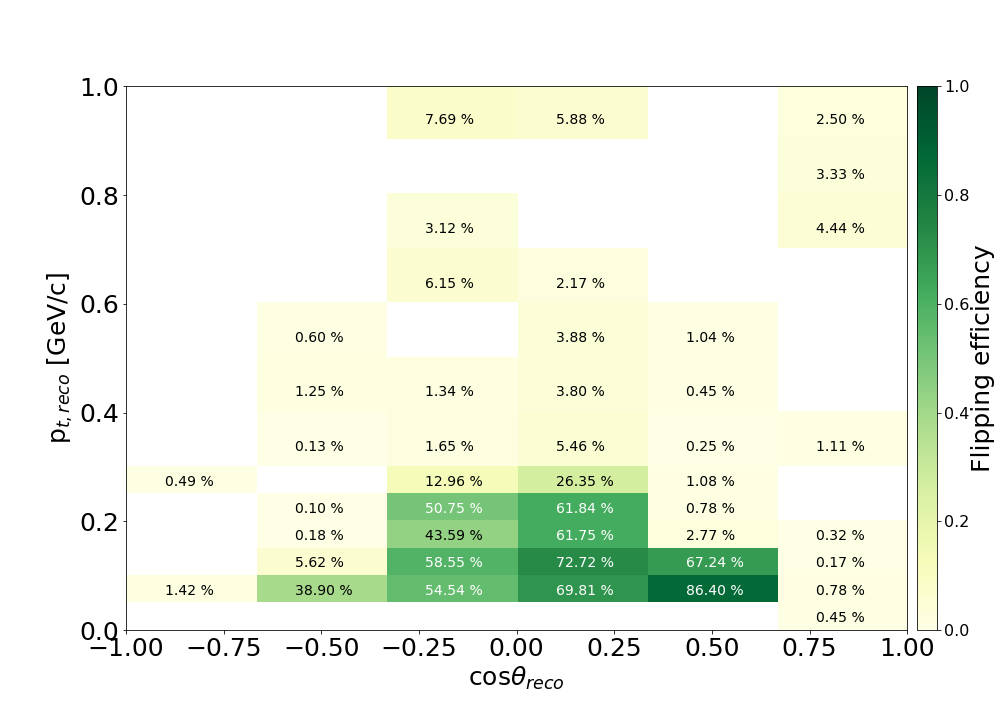}
    \label{fig:FlippingEfficiency}
  }
  \caption{Performance of the BDT based track flipping algorithm.
           The majority of tracks that are flipped had the wrong charge assigned to them previously \ref{fig:CorrectlyFlipped}, while up to 86.4\% of tracks with a previously wrong charge are flipped by the algorithm \ref{fig:FlippingEfficiency}.
          }
  \label{fig:Flipping}
\end{figure}

Recently, an asymmetry in the number of reconstructed positive and negative tracks was observed in data and later confirmed in simulation.
Subsequent studies revealed that this asymmetry originated, at least in part, from the mis-assignment of the charge: a small fraction of tracks was found correctly in terms of their hit content, but the wrong charge was assigned by the pattern recognition algorithms, and the track fit was not able to fix the mistake.
This effect mostly occurred in a specific region of the phase space, namely low $p_t$ curling tracks.
To reduce this mis-assignment, a method involving two multivariate methods, namely boosted decision trees (BDT) was developed.
The first BDT is trained to select tracks that potentially are assigned a wrong charge.
Input features include the seed information from the pattern recognition track (momentum and position of the first hit) as well as the hit content in SVD and CDC, and the track time.
All tracks selected by the first BDT are flipped in direction and charge sign and refitted using the GenFit2 DAF.
A second BDT is trained to determine whether the original or the flipped track is to be retained.
The second BDT uses the fit information from both the original and the flipped track, as well as the output of the first BDT.
Finally, the track list is updated by replacing the selected tracks with their flipped counterpart.
This approach results in 1.2\% of all tracks in generic $B\bar B$ events is selected by the first BDT and refitted, and 50\% of tracks with initially the wrong charge being flipped, effectively reducing the number of tracks with the wrong charge assigned to them by 50\% based on a study with 300000 simulated $B\bar B$ events.
Figure \ref{fig:CorrectlyFlipped} shows the fraction of tracks that were correctly selected by the first BDT and then flipped if they had a wrong charge assigned to them after pattern recognition.
The majority of the targeted tracks were successfully identified by the first BDT.
Figure \ref{fig:FlippingEfficiency} shows the efficiency of flipping tracks with a wrong charge assigned to them after pattern recognition.
Tracks in the forward region with a small transverse momentum and a wrong charge assigned to them are flipped with 86.4\% efficiency, while in the central phase space region up to 72\% of tracks are correctly flipped.

\section{Conclusions}

The track finding algorithms in Belle II are found to work reliably and with high efficiency both online and offline and are a key for Belle II to deliver world leading physics results in the early stage of the experiment.
In the future, precise timing information from the strip detector SVD will be used to calculate the EventT0 and to significantly reduce the number of fake tracks from non-triggered collisions by employing cluster time grouping.
A newly developed method to flip tracks and refit them using machine learning algorithms to reduce the number of tracks with wrong charge proves to work efficiently, reducing the number of tracks with wrong charge by up to 50\% in certain phase space regions.
The positive effect of both methods was shown in simulation and will be applied on data in the upcoming data taking period starting early 2024.
Further improvements to the track finding algorithms as well as further track refining steps will ensure that Belle II can produce world leading physics results for years to come.


\Acknowledgements

We thank the KEK and DESY computing groups for valuable support. We rely on many open-source software packages and thank the communities providing them. The author acknowledges the support from the Helmholtz-Gemeinschaft Deutscher Forschungszentren (HGF, Germany).



\end{document}

%% file: econfmacros.tex
\def\beq{\begin{equation}}
\def\eeq#1{\label{#1}\end{equation}}
\def\eeqn{\end{equation}}

\def\beqa{\begin{eqnarray}}
\def\eeqa#1{\label{#1}\end{eqnarray}}
\def\eeqan{\end{eqnarray}}

\def\Dslash{\not{\hbox{\kern-4pt $D$}}}
\def\dslash{\not{\hbox{\kern-2pt $\del$}}}

\def\msb{{\bar{\ssstyle M \kern -1pt S}}}

\def\s#1{\widetilde{#1}}
